\documentclass[sigconf]{acmart}
\usepackage[T1]{fontenc}
\usepackage{soul}
\AtBeginDocument{%
  \providecommand\BibTeX{{%
    \normalfont B\kern-0.5em{\scshape i\kern-0.25em b}\kern-0.8em\TeX}}}

\newcommand{\XXX}{{\sc Chatin }} 
\newcommand{\XXXdef }{{\sc Chat}bot conversat{\sc I}onal data exploratio{\sc N} environment } 

\begin{document}

\title[{Conversational Data Exploration}]
       {Conversational Data Exploration: \\ A Game-Changer for Designing Data Science Pipelines}

\author{Genoveva Vargas-Solar}
 \affiliation{%
  \institution{CNRS, Univ Lyon, INSA Lyon, UCBL, LIRIS, UMR5205, 69622}
   \city{Villeurbanne}
  \country{France}}
\email{genoveva.vargas-solar@cnrs.fr}
\author{Tania Cerquitelli}
\affiliation{%
  \institution{Department of Control and Computer Engineering, Politecnico di Torino}
  \city{Turin}
  \country{Italy}
}
\email{tania.cerquitelli@polito.it}

\author{Javier A. Espinosa-Oviedo}
 \affiliation{%
  \institution{CPE, CNRS, Univ Lyon, INSA Lyon, UCBL, LIRIS, UMR5205, 69622}
   \city{Villeurbanne}
  \country{France}}
\email{javier.espinosa@liris.cnrs.fr}

\author{François Cheval}
\affiliation{%
  \institution{CPE Lyon}
  \city{Villeurbanne}
  \country{France}
}
\email{francois.cheval@cpe.fr}

\author{Anthelme Bucaiile}
\affiliation{%
  \institution{CPE Lyon}
  \city{Villeurbanne}
  \country{France}
}
\email{anthelme.buchaille@cpe.fr}

\author{Luca Polgar}
\affiliation{%
  \institution{CPE Lyon}
  \city{Villeurbanne}
  \country{France}
}
\email{luca.polgar@cpe.fr}

\renewcommand{\shortauthors}{G. Vargas-Solar, et al.}

\begin{abstract}
This paper proposes a conversational approach implemented by the system \XXX for driving an intuitive data  exploration experience.
Our work aims to unlock the full potential of data analytics and artificial intelligence with a new generation of data science solutions. \XXX is a cutting-edge tool that democratises access to AI driven solutions, empowering non-technical users from various disciplines to explore data and extract knowledge from it. 
\end{abstract}




\maketitle

\section{Introduction}
Cutting-edge data science (DS) techniques have 
proven to be an alternative quantitative method for answering analytical and predictive research questions from non-data science experts. The principle is to process data collections containing observations of  phenomena using mathematical models, statistics, and artificial intelligence algorithms.
Non-expert users often find it challenging to effectively use data science methods, requiring significant interaction between data scientists and domain experts. Due to differences in language and expertise, mediators are needed to facilitate communication, which can be time-consuming and effort-intensive. To bridge this skills gap, companies can train domain experts in data science, train data scientists in the subject matter, or create a new Mediator role. A potential solution is to use a conversational AI mediator to fill the skill gap.

We argue that conversational approaches are a good way of (i) mediating the expression of data exploration requirements in natural language, (ii) transforming them into artificial intelligence-based analytics, and (iii) providing explanatory human understandable results.
Non-DS experts can compare results, give feedback to the system, and explore different analytics branches until satisfied. By learning from feedback and experience, the conversational environment can adapt its process to users' profiles.

{This paper introduces} \XXX (\XXXdef ) that proposes a conversational interface for creating data exploration interactive sessions for non-DS experts. The environment extends {ADESCA}\footnote{ADESCA implements interactive data exploration with automatically configured algorithms and graphical results.}\cite{bethaz2021enhancing} with conversational pipelines for creating broad and comparative data collection quantitative studies completed with the return of experiences that can improve the behaviour of the conversation sessions and the data exploration experience.
{To showcase} our vision, we use a case related to modelling gender biases in social norms. Driving the conversational exploration of a dataset released by the United Nations, the goal is to understand gender inequality across different countries.  We focus on the possibility of driving an intuitive  data exploration conversational task.

 Accordingly, the remainder of the paper is organised as follows. Section \ref{sec:relatedwork} synthesises existing work addressing conversational data exploration. Section \ref{sec:approach} introduces our conversational approach and system \XXX. 
 Section \ref{sec:usecase} illustrates  the validation of our approach and system through a use case 
 Section \ref{sec:conclusion} concludes the paper and discusses future work.

\section{Related work} \label{sec:relatedwork}
Chatbots \cite{adamopoulou2020chatbots,nagarhalli2020review} have become popular conversational interfaces for assisting humans in various tasks.
%
%
However, there is limited research on analytical chatbots that facilitate data science conversations using (semi) structured data.
%
We consider two approaches and solutions reported in the literature: \emph{conversational data exploration} and \emph{conversational interfaces} designed to provide ad-hoc services (e.g., searching for content). \\
\noindent
{\bf\em Conversational data exploration.}
N. Castaldo et al. \cite{castaldo2019conversational} propose a framework  for designing and generating chatbots to explore data. This approach utilizes data sources and modelling abstractions to enable designers to define the role of 
{critical  elements} in user requests.
 Articulate2 \cite{aurisano2016articulate2} is a conversational interface for visual data exploration that accepts natural language inputs. It enables generating and modifying visualisations while recording the user’s requests, visualisations, and findings.
%
{Analyza} \cite{dhamdhere2017analyza}  enables users to explore data through a question-and-answer feature in a spreadsheet product. It allows them to access a revenue/inventory database for a large sales force.
%
{V. Setlur et al.} \cite{setlur2022you} examined the Gricean Maxims  to design conversational interactions for data exploration, to support ambiguity and  handle intent.
 Hossein Hassan and Emmanuel Sirmal Silva \cite{hassani2023role} explore the role of ChatGPT in assisting data scientists with automating data cleaning and preprocessing, model training, and result interpretation. They assert that ChatGPT can aid with natural language processing tasks in data science, such as language translation, sentiment analysis, and text classification.
 {Code Interpreter} is a ChatGPT 4 plugin\footnote{\url{https://www.datanami.com/2023/07/11/openai-releases-chatgpt-code-interpreter-your-personal-data-analyst/}} that allows users to upload and analyze files using natural language (e.g., converting files, performing mathematical operations, plotting data). The Code Interpreter generates Python code that is then executed in a sandboxed environment that lasts the conversation's duration. The execution result (e.g. a plot) is then incorporated into the ChatGPT conversation with an explanation of what was done.
 \ul{We believe current conversational solutions offer robust tools for creating dialogues that enable non-data science specialists to design quantitative analytics. 
 Our work aims to improve conversational solutions by suggesting proactive actions to guide non-specialists in intuitively designing quantitative analytics and exploring data to extract insights and value from it.}
 
\noindent
{\bf\em Conversational bots interface building tools.}
Dialogflow\footnote{\url{https://cloud.google.com/dialogflow/docs}} is a natural language understanding platform created by Google that simplifies the process of designing and integrating a conversational user interface into mobile or web applications. 
Amazon Lex is a service provided by Amazon Web Services (AWS)\footnote{\url{https://docs.aws.amazon.com/lex/latest/dg/what-is.html}} that allows developers to embed conversational interfaces into voice and text applications. It uses natural language understanding (NLU) and automatic speech recognition (ASR) to create lifelike and engaging user interactions. With Amazon Lex, you can create chatbots, virtual assistants, and other conversational experiences that feel natural and intuitive. 
ParlAI \cite{miller2017parlai} is a platform designed by the Facebook AI Research (FAIR) team to facilitate collaboration and reuse of significant dialogue tasks among researchers. It encourages collaboration in developing NLP systems, including integration with bots and humans. This community-based platform allows researchers to work together to advance the field of natural language processing and improve conversational AI.
Language Understanding (LUIS)\footnote{\url{https://docs.microsoft.com/en-us/azure/cognitive-services/luis/what-is-luis}} is a cloud-based conversational AI service developed by Microsoft. It is designed to extract valuable information from dialogues by interpreting user goals (intents) and distilling useful information from sentences (entities). LUIS uses a high-quality language model to achieve this, and it integrates seamlessly with the Azure Bot Service, making it easy to create sophisticated bots that can understand and respond to natural language input. 
\ul{Our work will use chatbot implementation tools to design and implement a conversational interface to guide non-DS specialists through conversations to explore data collections and produce and explain quantitative insights.}\\
\noindent
{\em Discussion.}
Using conversational assistants and interfaces for querying and processing data is not new. Approaches like \emph{query-by-example} \cite{zloof1977query} in the early years of relational querying, natural language-based database querying \cite{Hozcan2020state}, and visual query languages \cite{bhowmick2022data} are examples of previous solutions. 
Conversational interaction between users and systems  includes works that address the SQL-to-NL (Natural Language) and NL-to-SQL problems such as \cite{Eleftherakis2021LetTD,weir2019dbpal,brandt2020towards,brunner2021valuenet}. However, these approaches are primarily focused on query-oriented interactions. Other approaches, such as \cite{kumar2016towards}, explore dialogue-based approaches for supporting rich data visualisations. 

In recent years, there has been a growing trend towards using high-level visual specifications in Machine Learning (ML) studios, such as Microsoft ML Studio, to simplify the expression and programming of data science and analytics pipelines. However, even with these tools, technical and mathematical expertise is still required. To address this, new strategies are emerging that use natural language processing (NLP) interfaces to allow users to express their requirements more intuitively and user-friendly. For example, English can be used as a programming language for Apache Spark \footnote{\url{https://www.databricks.com/blog/introducing-english-new-programming-language-apache-spark}}. With the emergence of ChatGPT and GitHub Copilot, this process is becoming even more intuitive and exploratory, allowing users to express their high-level requirements more easily.
\ul{ We aim to provide a complete dialogue with the user, moving beyond simple query-and-answer interactions. Our goal is to guide users through data exploration, where they can  (1) intuitively use and calibrate statistical and machine learning tools to achieve their analytics targets; (2) and, in consequence, design DS pipelines.}

\section{\XXX: a conversational data exploration environment} \label{sec:approach}
\begin{figure*}[htbp]
\centerline{
\includegraphics[width=0.8\textwidth]{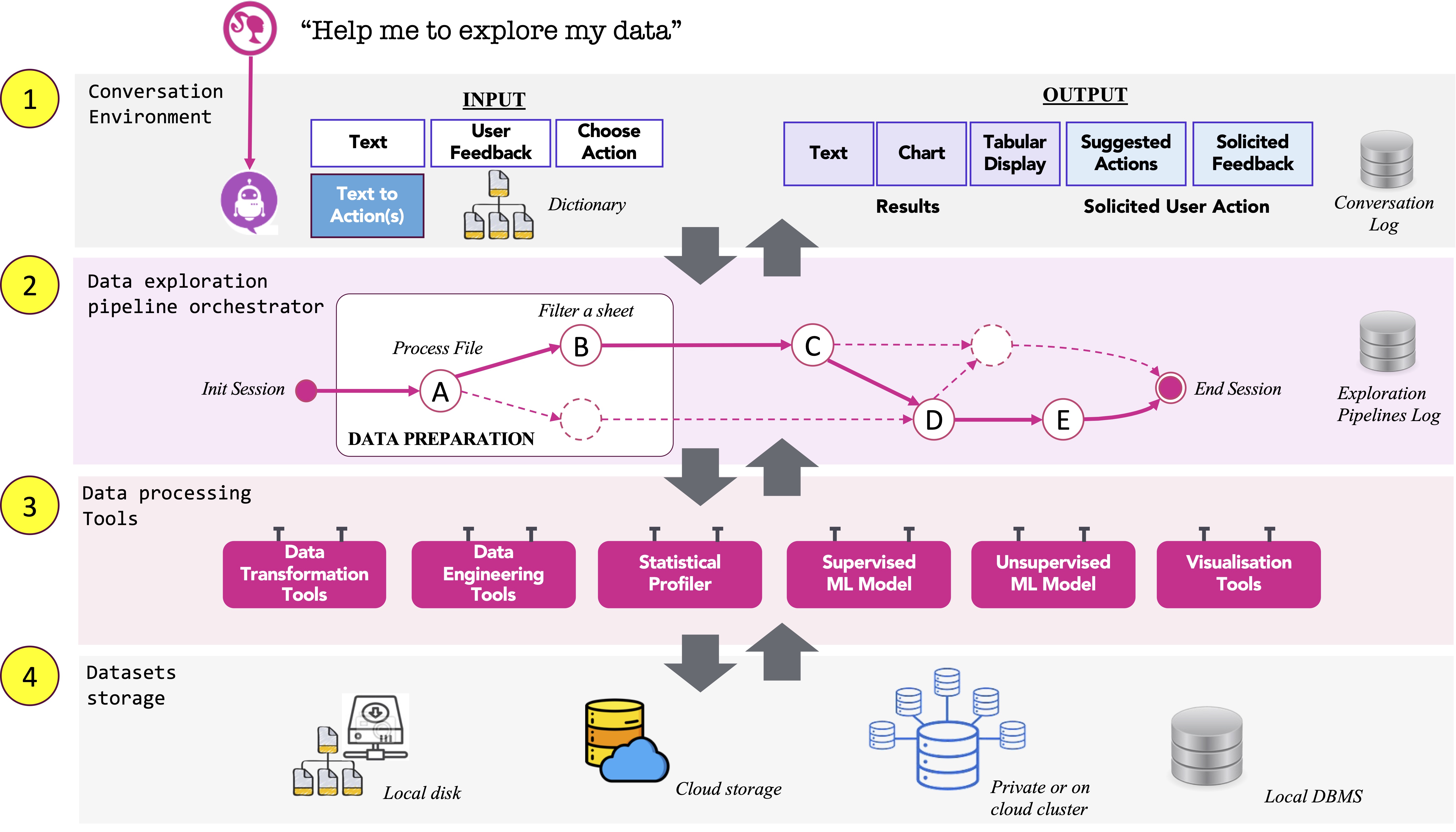}
}
\caption{\XXX general architecture}
\label{fig:architecture}
\end{figure*}

Figure \ref{fig:architecture} shows the general architecture of \XXX, a conversational environment designed to assist users in exploring data collections and intuitively defining data exploration pipelines. The architecture, which was first proposed in a position paper \cite{9671883}, is organised into four layers:
(1) The conversational layer, where the user interacts with the system using natural language requests and provides feedback on the system’s output.
(2) The orchestrator layer executes the data exploration pipeline.
(3) The data processing layer, which offers tools and APIs for data processing.
(4) The storage layer, which provides support for persistent data collections, exploration operation results, and conversations.

As shown in Figure \ref{fig:architecture}:1, the conversation environment provides an interface for the user to express their requirements in natural language. These requirements are then transformed into one or more data processing tasks or interactions that request additional input. The conversation relies on meta-patterns as a reference for performing quantitative exploration tasks. The actual data exploration, data processing (such as clustering models), assessment scores, and visualisation tools are implemented by a backend engine that serves as a toolkit developed in previous work \cite{bethaz2021enhancing} (see Figure \ref{fig:architecture}:3).

The conversational pipelines executed by the coordinator, as shown in Figure \ref{fig:architecture}:2, are performed within sessions. Requests are processed by utilizing logs from previous explorations conducted by the system, identifying similarities and connections between various analyses to present the most relevant and appropriate results for the user’s request. These results can be displayed in text, table, or graph format. The user can then assess the clarity and usefulness of the results by providing feedback, which is added to the exploration log and used to inform future analyses. 

This abstraction layer shields the user from the complexities associated with  data exploration and analysis libraries and tools, resulting in a more user-friendly experience.

\begin{figure}[htbp]
\centerline{
\includegraphics[width=0.37\textwidth]{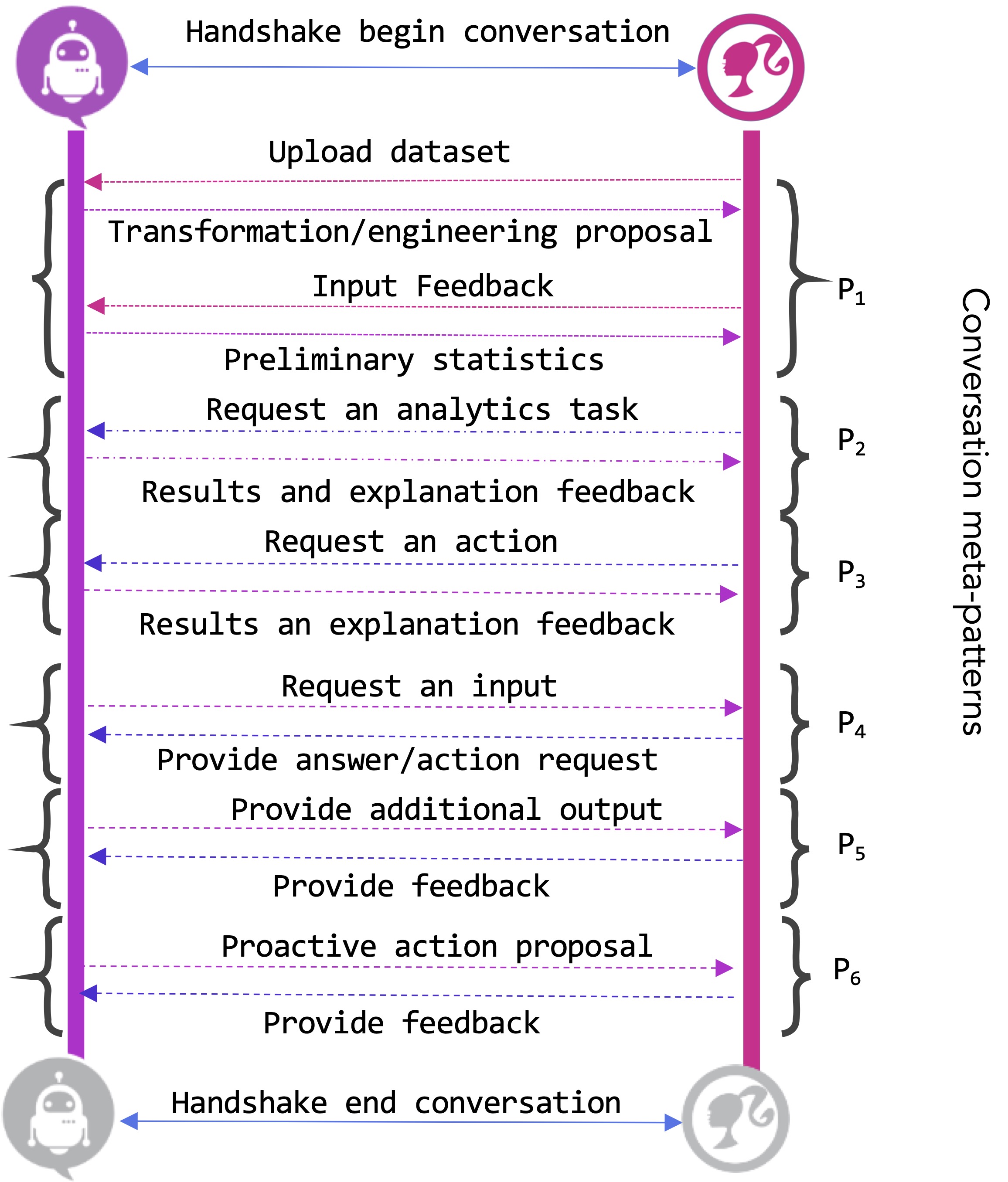}
}
\caption{Conversational meta-patterns}
\label{fig:conversation}
\end{figure}

\noindent
{\bf\em Conversational meta-patterns.}
The sequence diagram of the meta-conversation between \XXX and the user is shown in Figure \ref{fig:conversation}. It is a two-way interaction initiated and concluded by the user or the system. Every user request receives a response from the system. The system can also ask the user to do actions, for example, providing additional information or feedback.

An example of conversational interaction can start
with a dialogue where 
the user and the system interact to explore the dataset and produce initial statistical information. Instead of asking the user to know formats and remember formulae to compute statistical metrics, the system computes the quantitative profile directly and provides a textual explanation of the content.

\noindent
{\em Proactive strategy}: Represents a data preparation interaction.
Rather than just responding to a user request, \XXX integrates a proactive approach during the conversation to anticipate/suggest/analyse activities and queries that could interest the user. The proactive strategy is a novel and original feature of \XXX compared to existing conversational chatbots.

We have identified five types of interactions in Figure \ref{fig:conversation}, involving   specific questions and answers: 
\\
\noindent
- Pattern P$_1$: Represents a {\em data preparation dialogue}. The conversation pattern starts with a request like "\emph{I need help exploring the following data collection}" or "\emph{I want to explore a data collection?}". Consequently, the system asks the user to upload the data collection and acknowledges the action by telling the user the file format. Then, it proposes a structural or statistical description of the data (e.g., "\emph{do you prefer a structural and/or a statistical description of your data?}"). \\
\noindent
- Pattern P$_2$: Represents an {\em analytics task request dialogue}. The user can request an analytical task, such as data modelling or prediction, while delegating the complexity of the analysis to the system. 
The system can then choose and propose a possible solution, performing the task and producing output with explanations of the task done. Often the results are associated with plot options so that the user decides which one s/he prefers. \\
\noindent
- Pattern P$_3$: Represent an {\em action request dialogue.}
The user can also request more specific actions, such as changing the time interval for analysis or excluding a particular attribute from future analysis. \\
\noindent
- Pattern P$_4$: Represents a {\em request input dialogue initiated by the chatbot.}
 The system can ask for additional input parameters in terms of interactions initiated by the system.  Additionally, the system may provide output to the user without being specifically requested. The output type is decided according to the user's profile and previous experience, which can help determine what analysis and results would be most relevant and relevant to the user without any initial input.\\
\noindent
- Pattern P$_5$: Represents a {\em an additional output provision dialogue initiated by the chatbot.}
Following the previous example of pattern P$_1$,
 the system  produces  various statistics from a dataset, and it can follow up by proposing plots with a narrative description of the type of plot and how it represents the data. For the plots, the chatbot suggests plots with explanations and asks for feedback from the user so that she/he can choose the plots that seem more comprehensible.\\
\noindent
- Pattern P$_6$: 
Represents a \textit{proactive interaction initiated by the chatbot.} The system can start a conversation and suggest activities and mini-tasks based on the system skills and expertise acquired from past interactions with users and users' provided feedback. The user can accept or reject the suggestion. \\
%
%
%
The conversational meta-patterns can occur in any order.  The conversation can also be paused anytime, allowing the user to save the progress and resume later without losing any analysis results.

\noindent
{\bf\em  Data exploration assistance: converting text to action.}
Once \XXX receives a request, it  divides it into tasks ordered in an execution plan. 
The execution plan is a directed graph representing a  solution space with data exploration functions that can be requested according to the initial handshake. It represents a dataflow with  operations represented by nodes and edges representing the input/output data. 

The execution of a node implies resolving it to a concrete implementation in a library or provided by an external module (e.g., transforming the dataset into some format, performing statistical profiling, or clustering values and producing plots). Every task in the execution plan implements an interactive cycle that receives input, proposes/asks for input parameters, presents results, and expects feedback. 

\begin{figure}[htbp]
\centerline{
\includegraphics[width=0.40\textwidth]{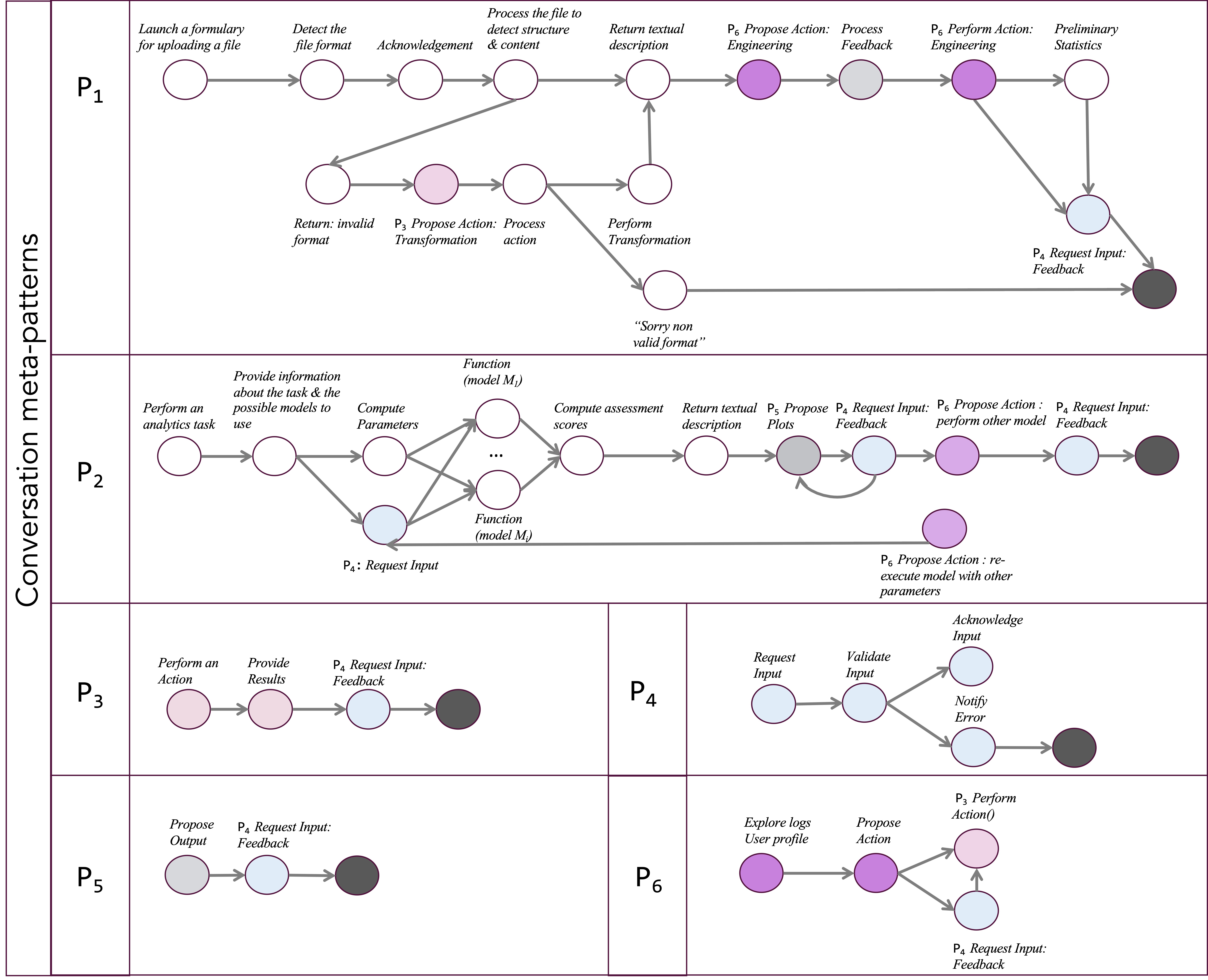}
}
\caption{Transformation conversation meta-patterns towards execution plan patterns.}
\label{fig:transformation-rules}
\end{figure}

\noindent
{\em  Text to action strategies: building data exploration execution plans.}
We have adopted two strategies to implement the text-to-action strategy of the system.\\
\noindent
- {\sc\small Simplified text-to-action strategy:} 
The general principle of the strategy relies on a set of rules and a controlled vocabulary extracted from Wordnet that contains the verbs associated with data exploration actions (compute, analyse, transform, search, give, etc.) and nouns like data, dataset, data collection, table, etc. The extracted vocabulary also contains synonyms to enable fluent interaction. The sentences the user issues are tokenised to search for keywords in particular verbs and nouns. Data exploration actions are associated with one or several potential conversational patterns. 

Every conversational pattern has a corresponding execution plan graph pattern (see Figure \ref{fig:transformation-rules}). 
As shown in the figure, the data transformation and preparation request described by pattern P$_1$,  is transformed into an execution plan represented by a directed acyclic graph. 
A node in the plan can correspond to an abstract exploration function,
an action request of the bot to the user and an action from the user. 

\noindent
-  {\sc\small Second strategy} relies on using an existing chatbot API to deal with complex conversations. The interpreted actions are then used for building the execution plan that then guides the conversational exploration, guiding the user to accomplish this task.
This strategy is undergoing, and the objective is to train the chat model with conversations to produce execution plans. The execution of the plans with the chatbot ensuring the conversational interaction can then assist users in exploring data collections intuitively. 

\begin{figure}[htbp]
\centerline{
\includegraphics[width=0.40\textwidth]{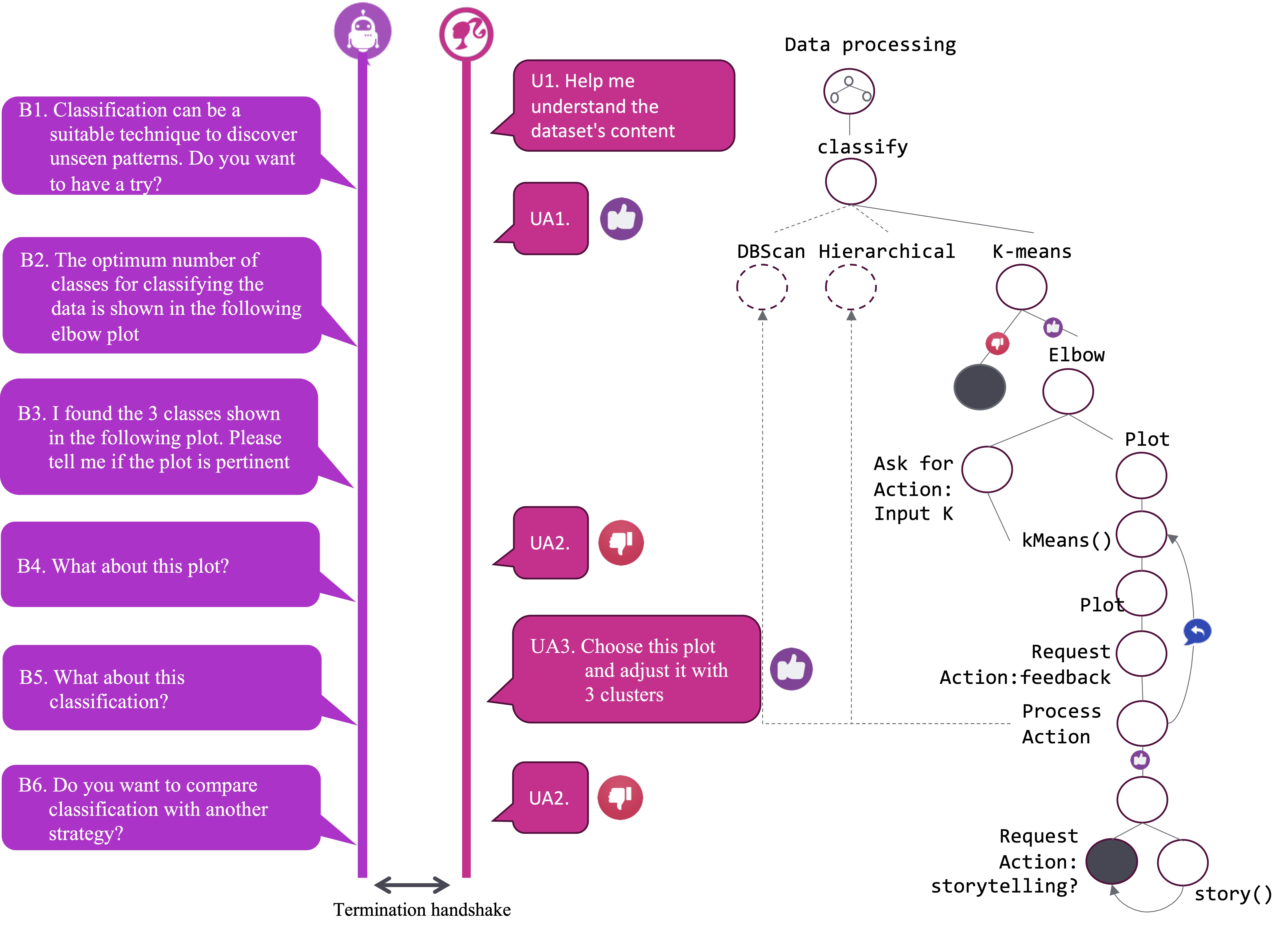}
}
\caption{Execution plan of a conversational data exploration pattern.}
\label{fig:execution-plan}
\end{figure}

 To go beyond the exploration, a user would like to extract further understanding of the data collection content: {\em "I want to understand the dataset's content further"}. The bot proposes to compare different classification techniques \cite{soofi2017classification}:
 {\em "Classification can be a suitable technique to discover unseen patterns in your data collection. I can perform different strategies. Do you want to have a try?"} The execution plan shown on the right side in figure \ref{fig:execution-plan} illustrates the non-resolved execution plan. A first try can be to choose the K-means algorithm \cite{faber1994clustering}. This choice launches several conversation patterns. The dotted nodes represent potential tasks to be executed if the user selects to test further classification techniques (e.g., SVM \cite{yue2003svm}).

 If the conversation is paused, the system stores the resolved plan with the execution state of every task. Then, it can be retrieved and restarted as the conversation is resumed. When the session is terminated, the system has the execution log of the conversation and the exploration pipeline designed by the user (intuitively).

\section{Exploring gender biases due to social norms} \label{sec:usecase}
Our use case is inspired by the Human Development Report produced by the United Nations Development Programme\footnote{\url{https://hdr.undp.org}}. The report  proposed a Gender Social Norm Index (GSNI) to measure public attitudes towards women across four dimensions (political, educational, economic, and physical integrity) to assess how biased beliefs can impact gender equality and human rights \cite{undp2023}.
The intention is to explore a dataset released in 2023 through conversational exploration to understand its content. The analysis aims to compute metrics for \textbf{many} dimensions across different countries and identify correlations of biases that impact gender inequality \textbf{as well as biases similarity across different countries}. The focus is on the conversational aspects and their connection with data processing.

\noindent
{\bf\em World Values Survey (WVS) Dataset }
 assesses social, political, economic, and cultural values and norms through questionnaires. The dataset contains microdata captured in different waves and examines shifts in cultural values, attitudes, and beliefs across countries and over time. The use case uses the latest  dataset contained in a 136.27 KB Excel file \footnote{\url{https://hdr.undp.org/sites/default/files/publications/additional-files/2023-06/GSNI2023_AnnexTables2.xlsx} produced on 12 January 2023.}.

\begin{figure*}[htbp]
\centerline{
\includegraphics[width=0.85\textwidth]{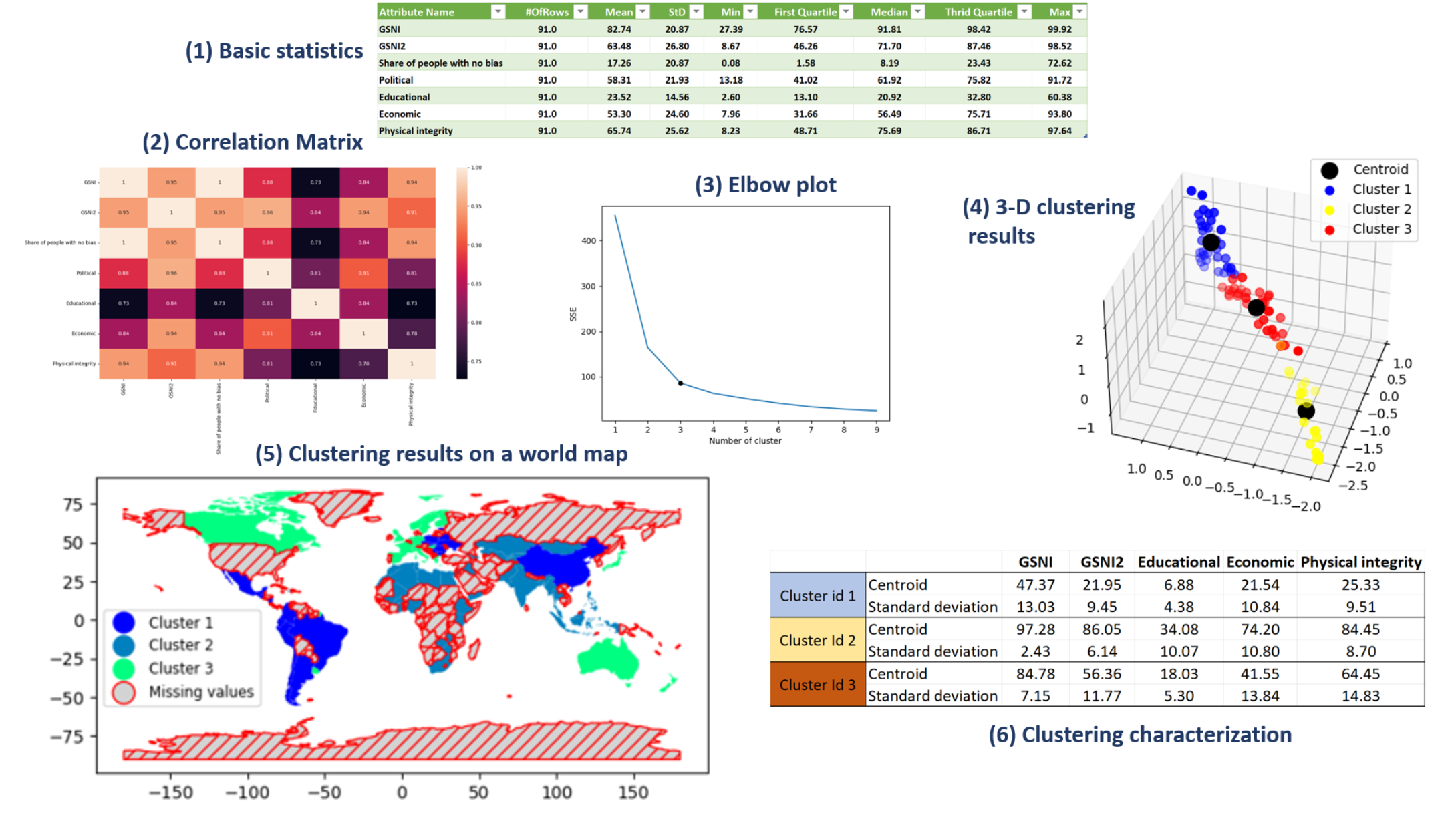}
}
\caption{\XXX: A storytelling for the WVS Dataset.}
\label{fig:Chatin_Storytelling}
\end{figure*}

\noindent
{\bf\em Exploration conversations.}
This use case allowed for testing several exploration patterns.\\
c
 Data preparation ( transformation, feature engineering, and quantitative profile). The GSNI2023 file was uploaded in response to a user handshake request:
{\sc\small U1. }{\em "Help me to analyse my data"}. The request launches  an instance of the conversation pattern P$_1$ shown in  figure \ref{fig:execution-plan} (BX refers to a request produced by the bot, UAX refers to an action done by the user, UX tags a user request):  \\
  \noindent
    {\sc\footnotesize --- B1. } {\footnotesize\em "Please upload your data collection."} 
   --- {\sc\footnotesize UA1. } {\footnotesize\em Upload file}\\
    \noindent
    {\sc\footnotesize --- B2. } {\footnotesize\em  “Your collection is an Excel file with 5 sheets 
    It must be transformed into CSV. Do you want me to transform all sheets?”} 
    {\sc\footnotesize --- U2. }{\footnotesize\em "Transform only the first sheet."}\\
    \noindent
  {\sc\footnotesize 3B.} {\footnotesize\em “We are going to start a data transformation process”} \\
  {\footnotesize \XXX produces a transformed CSV file and a basic statistical description (see the results produced by our system in the top side of Figure \ref{fig:Chatin_Storytelling}:(1))}.\\
    \noindent
    {\sc\footnotesize --- U3. } {\footnotesize\em "Analyse the linear correlation between each couple of numerical attributes in the dataset".} \\
    {\footnotesize \XXX produces a correlation matrix, showing the Pearson coefficient computed for each couple of attributes as shown in Figure \ref{fig:Chatin_Storytelling}:2}.\\
     {\sc\footnotesize  ---- B4.} {\footnotesize\em "Are these results what you expected?"} ---  {\sc\footnotesize  ---- U4.} {\footnotesize\em "Ok."}\\
\noindent
    The final handshake includes the user's feedback and a goodbye message from the bot.

\noindent
- [Test 2 composing meta-patterns  P$_2$, P$_3$, and P$_5$:]
  The conversation starts with the user request {\em "I want to understand the data further."} (meta-pattern P$_2$). The chatbot proposes an explanation of the role and principle of classification by the bot (meta-pattern P$_5$). It suggests a first initiative (with k-means, without telling the user) and launches a conversation for collecting some data that serve as input to calibrate the analytics task. For example, the chatbot says {\em B1: "According to the characteristics of your data, I have selected the numerical attributes. I have  first removed  correlated  attributes. Then I automatically determined the optimum number of groups to classify the data.  Based on the analysis, I should organise your data into 3 clusters. Here are the results."} The \XXX produces  the elbow plot and a 3-d representation of the clustering with the corresponding centroids.  According to meta-pattern P$_3$, the chatbot says {\em B2: "Do you want me to explain the notion of attributes correlation and how it is computed?"}. On acceptance of the user, the chatbot explains: {\em "Variable correlation is determined by computing the Pearson coefficient. I decided to eliminate variables with a coefficient greater than 0,95."} (meta-pattern P$_5$).

\noindent
- [Test 3 Proactive action P$_6$:]
The chatbot proactively generates visualisation options with explanations so the user can provide feedback and ask to adjust the result or try a different strategy. To conclude the session, the system proposes a narrative (storytelling) of the whole conversation with plots intended to allow a better understanding of the data. We do not show the narrative here, but Figure \ref{fig:Chatin_Storytelling} shows the plots it contains.
The plot in Figure \ref{fig:Chatin_Storytelling}:5 shows clustering results on a world map, highlighting groups of countries with similar biases. Because the dataset provides data for only a subset of countries, the map shows countries in the world map for which data are unavailable. To give more insight into the clustering results, Figure \ref{fig:Chatin_Storytelling}:6 shows each group's centroids (in terms of average and standard deviation). The user can see that the clustering provided consists of coherent and well-separated groups of countries regarding the bias indices. As shown in Figure \ref{fig:Chatin_Storytelling}:(6), each group has a centroid characterised by values for all input data (i.e., well-separated groups) that are very different from the values assumed by the same input for the other centroids, while the standard deviation for each input variable is often very low, showing that the identified groups are very cohesive.



\noindent
{\bf\em Assessment setting.}
One of our contributions is proposing a set of assessment metrics for our approach that can go beyond interpreting user requests. Our work is not willing to compete with existing very powerful chatbots. Our contribution concerns the design of an intuitive data exploration process where the guidance comes from the system and not the user, a priori non-DS specialist.
We propose the following assessment metrics: \\
    \noindent
    [M1] The number of user-chatbot interactions the system requires for identifying the user's intention modelled by a meta-pattern.\\
    \noindent
    [M2] Precision and recall of the execution plans.\\
    \noindent
    [M3] Acceptability of the responses given by the number of liked feedback. \\
    \noindent
    [M4] Data exploration task completion given by the number of sessions that lead to the generation of storytelling.\\
    \noindent
    [M5] {Customer satisfaction metrics, such as \textit{Customer satisfaction score} by asking customers to rate their experience after an interaction with \XXX. \textit{Net promoter score} (it measures if customers would be willing to recommend \XXX\ to other people (Customers who give a score of 9 or 10 are called promoters, from o tp 6 detractive and from 7 to 8 passive)}. \\
\noindent
{\bf\em Scope and Limitations.}
We have tested our approach under a proof of concept introduced in this section. We have also run tests exploring datasets from our previous projects \cite{cerquitelli2020data,bethaz2021empowering}. However, the users have advanced DS knowledge even if they have other domain expertise. This experience has allowed defining assessment metrics.
Therefore, we are currently setting two types of tests: (1) generating synthetical conversations for testing our system for metrics M1 and M2; (2) organising groups of users with absent, low and medium IT expertise to test M3, M4 and M5.

We envisioned a data exploration pipeline design where interactive, conversational approaches empower users of all backgrounds and expertise levels to explore data. Conversations must be agile, intuitive, and user-friendly to achieve this goal. These conversations should be tailored to the data and the user’s profile, guiding the user toward a clear understanding of the data and the exploration tools available for their task.

\section{Conclusion and Future Work} \label{sec:conclusion}
This paper introduced \XXX, a conversational data exploration system allowing users to design their pipelines. The system's objective is to propose an explorative and interactive design environment where users can understand the content of data collections and produce a quantitative profile of their content going back and forth in their analytics tasks. Once the exploration converges, the system aggregates and synthesises the process into a storytelling document.

Our future work aims to make the conversation flow more natural by interpreting user requests and mapping them into actions considering the data's nature. We will also provide personalised answers, narratives, and visualisations to deliver easily understandable data stories. Additionally, we are working on predicting and recommending the next task in the data exploration process based on previous actions and dataset characteristics.


\bibliographystyle{ACM-Reference-Format}
\bibliography{sample-base}


\end{document}